\documentclass[12pt]{iopart}

\usepackage[dvips]{graphicx}
\usepackage{array}
\usepackage{amssymb}
\usepackage{hhline}
\usepackage{longtable}
\usepackage{dcolumn}% Align table columns on decimal point
\usepackage{bm}% bold math
\usepackage{subfigure}
\usepackage[usenames]{color}
\begin{document}

\title[Mean first-passage time for random walks on the T-graph]{Mean first-passage time for random walks on the T-graph}

\author{Zhongzhi Zhang$^1$${}^,$$^2$, Yuan Lin$^1$${}^,$$^2$, Shuigeng Zhou$^1$${}^,$$^2$, Bin Wu$^1$${}^,$$^2$, Jihong Guan$^3$}

\address{$^1$ School of Computer Science, Fudan
University, Shanghai 200433, China}
\address{$^2$ Shanghai Key Lab of
Intelligent Information Processing, Fudan University, Shanghai
200433, China}
\address{$^3$ Department of Computer Science and Technology,
Tongji University, 4800 Cao'an Road, Shanghai 201804, China}
 \eads{\mailto{zhangzz@fudan.edu.cn}, \mailto{sgzhou@fudan.edu.cn}, \mailto{jhguan@tongji.edu.cn}}

\begin{abstract}
For random walks on networks (graphs), it is a theoretical challenge
to explicitly determine the mean first-passage time (MFPT) between
two nodes averaged over all pairs. In this paper, we study the MFPT
of random walks in the famous T-graph, linking this important
quantity to the resistance distance in electronic networks. We
obtain an exact formula for the MFPT that is confirmed by extensive
numerical calculations. The interesting quantity is derived through
the recurrence relations resulting from the self-similar structure
of the T-graph. The obtained closed-form expression exhibits that
the MFPT approximately increases as a power-law function of the
number of nodes, with the exponent lying between 1 and 2. Our
research may shed light on the deeper understanding of random walks
on the T-graph.
\end{abstract}

\pacs{05.40.Fb, 89.75.Hc, 05.60.Cd, 05.10.-a}

%87.23.Ge Dynamics of social systems
%87.23.Kg Dynamics of evolution
%89.75.Hc Networks and genealogical trees
%02.50.Le Decision theory and game theory
%02.10.Ox Combinatorics; graph theory

%\pacs{05.40.Fb}{Random walks and Levy flights}
%\pacs{89.75.Hc}{Networks and genealogical trees}
%\pacs{05.60.Cd}{Classical transport}
%\pacs{05.10.-a}{Computational methods in statistical physics and nonlinear dynamics}
%Uncomment for PACS numbers title message
%\pacs{00.00, 20.00, 42.10}
% Keywords required only for MST, PB, PMB, PM, JOA, JOB?
%\vspace{2pc}
%\noindent{\it Keywords}: Article preparation, IOP journals
% Uncomment for Submitted to journal title message
%\submitto{\JPA}
% Comment out if separate title page not required
\maketitle
%%%%%%%%%%%%%%%%%%%%%%%%%%%%%%%%%%%%%%%%%%%%%%%%%%%%%%%%%%%%%%%%%%%%

\section{Introduction}

As the paradigmatic discrete-time realization of Brownian motion and
diffusive processes, whose theory was formulated in the pioneering
work by Einstein~\cite{Ei1906} and Smoluchowski~\cite{Sm1906}
published in beginning of the last century, random walks have
received a considerable amount of
attention~\cite{Sp1964,We1994,Hu1996}, uncovering their wide range
of distinct applications. Thus far, random walks are still an active
area of research~\cite{MeKl00,BeHa00,HaBe02,MeKl04,BuCa05}. However,
due to the complexity and variety of real media, theory of random
walks on general graphs (networks) is not yet available. For this
reason, studying random walks occurring on simple structures is a
matter of exceptional importance. Fractal structures, in particular
deterministic fractals, are valuable media in this
content~\cite{BeHa00,HaBe02}, because their properties can be
exactly studied.

Among various deterministic
fractals~\cite{Ma82,Vi83,ZhZhChYiGu08,BeOs79,GrKa82,
ZhZhFaGuZh07,ZhZhZoCh08}, the T-fractal (T-graph)~\cite{KaRe86} is a
typical candidate for exactly solvable model, and a plethora of
issues for random walks on this fractal have been
studied~\cite{KaRe89,Ma89,MaSaSt93,GiMaNat94,KnKn96,MaHa89,BeMeTeVo08,HaRo08,Ag08}.
A simple analytical approach was proposed in~\cite{KaRe86}, showing
that random walks on the T-graph can be mapped onto diffusion on a
comb structure. The asymptotic behavior of the moments of the
first-passage time and survival probability for random walks on the
T-graph was computed in~\cite{KaRe89}. Particularly, random walks
performed on the T-graph with a single trap were extensively studied
by many groups~\cite{KaRe89,MaHa89,HaRo08,Ag08}, revealing some
scaling relations and dominating behavior. Despite the fact that
these investigations uncovered many unusual and exotic features of
the T-graph, providing useful insight into understanding random
walks on this fractal, they did not give a complete picture of
random-walk dynamics on  the T-fractal, since in these studies, only
one special trap was considered. It was shown that in some real
networks, any node may be looked as a trap, and the location of
traps strongly affects the behavior of random
walks~\cite{KiCaHaAr08}. Irrespective of its obvious importance and
ubiquity, this issue has not been addressed for the T-graph.

In this paper, using the connection between the random walks and
electrical networks~\cite{DoSn84}, we investigate random walks on
the T-graph. The random walk process addressed here may be
considered as the trapping problem with the perfect trap uniformly
located at all nodes. We derive analytically an exact formula for
the mean first-passage time (MFPT) averaged all pairs of nodes,
which characterizes the efficiency of random walks on the T-graph.
We show that the location of trap has no qualitative effect on the
scaling for MFPT. We expect that our analytical method can be
applied to some other deterministic media, and that our results can
lead to deeper insights to random walks on the T-graph.

%%%%%%%%%%%%%%%%%%%%%%%%%%%%%%%%%%%%%%%%%%%%%%%%%%%%%%%%%
% Figure  1
%%%%%%%%%%%%%%%%%%%%%%%%%%%%%%%%%%%%%%%%%%%%%%%%%%%%%%%%%%
\begin{figure}
\begin{center}
\includegraphics[width=.30\linewidth,trim=100 30 100 0]{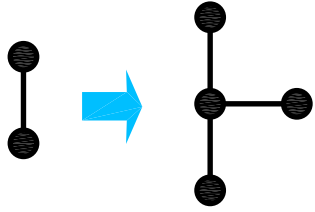}
\caption{ (Color online) Iterative construction method of the
T-graph. The next generation is obtained by performing the operation
shown in the right-hand side of the arrow.} \label{cons}
\end{center}
\end{figure}
%%%%%%%%%%%%%%%%%%%%%%%%%%%%%%%%%%%%%%%%%%%%%%%%%%%%%%%%%%

\section{Brief introduction to the T-graph}

\begin{figure}
\begin{center}
%\fbox{\includegraphics[width=.85\linewidth,trim=100 0 100 0]{Vicsek}}
\includegraphics[width=.75\linewidth,trim=130 0 100 0]{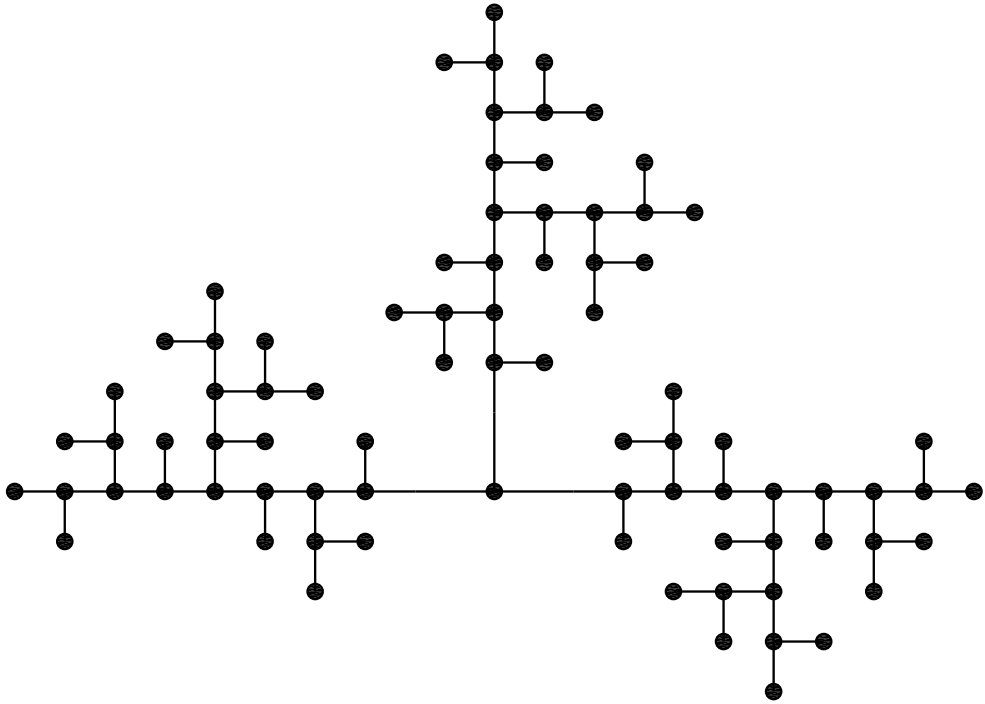}
\caption{Illustration of the T-graph of generation 4.} \label{net}
\end{center}
\end{figure}

The T-graph is built in an iterative way~\cite{KaRe86,Ag08}. We
denote by $\mathbb{T}_{g}$ ($g\geq 0$) the T-graph after $g$
iterations. Note that henceforth we will also call the number of
iterations as generation of the T-graph. The construction of the
T-graph starts from ($g=0$) two nodes connected by an edge, which
corresponds to $\mathbb{T}_{0}$. For $g \geq 1$, $\mathbb{T}_{g}$ is
obtained from $\mathbb{T}_{g-1}$ by performing the operation
illustrated in Fig.~\ref{cons}. According to the construction
algorithm, at each generation, the number of edges in the system
increases by a factor of $3$. Thus, we can easily obtain that at
generation $g$, the total number of edges in $\mathbb{T}_{g}$ is
$E_{g}=3^g$. Since the T-graph is a tree, the total number of nodes
in $\mathbb{T}_{g}$ is $N_{g}=E_{g}+1=3^g+1$. Figure~\ref{net} shows
schematically the structure of $\mathbb{T}_{4}$.

%%%%%%%%%%%%%%%%%%%%%%%%%%%%%%%%%%%%%%%%%%%%%%%%%%%%%%%%%
% Figure  2
%%%%%%%%%%%%%%%%%%%%%%%%%%%%%%%%%%%%%%%%%%%%%%%%%%%%%%%%%%
%%%%%%%%%%%%%%%%%%%%%%%%%%%%%%%%%%%%%%%%%%%%%%%%%%%%%%%%%%

The T-graph is a fractal with the fractal dimension and random walk
dimension being $d_f=\frac{\ln3}{\ln2}\approx 1.585$ and
$d_w=\frac{\ln 6}{\ln2}=1+d_f$, respectively. The spectral dimension
of the T-graph is $\tilde{d}=2\,d_f/d_w=\frac{\ln9}{\ln6}\approx
1.226$. Note that for a general connected graph, the spectral
dimension governs the longtime behavior of a random walk on the
graph at a deep level. For example, when a walker originating at a
given node $i$ of the graph, the probability $P_{ii}(t)$ for
returning back to $i$, at long time $t$, obeys the relation
$P_{ii}(t)\sim t^{-\tilde{d}/2}$~\cite{Ge82}. Again for instance,
for an infinite graph with the spectral dimension $\tilde{d} \leq
2$, a walker starting from a given node will return back to the node
almost surely over the course of time. The phenomenon of a random
walk on such a graph is called ``persistence". Since the spectral
dimension $\tilde{d}<2$ for the T-graph, a random walk on such a
graph is persistent.

\begin{figure}
\begin{center}
\includegraphics[width=0.20\linewidth,trim=100 0 100 0]{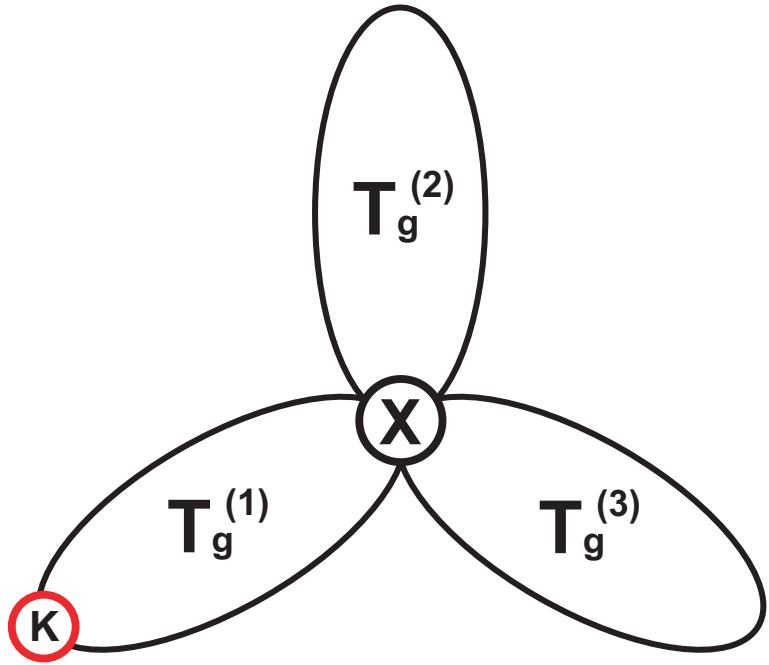}
\caption{(Color online) A schematic illustration of the second
construction method for $\mathbb{T}_{g+1}$, which is obtained by
joining three copies of $\mathbb{T}_{g}$, consecutively represented
as $\mathbb{T}_{g}^{(1)}$, $\mathbb{T}_{g}^{(2)}$, and
$\mathbb{T}_{g}^{(3)}$, respectively. } \label{copy}
\end{center}
\end{figure}

To facilitate the description in what follows, we define the central
node in Fig.~\ref{net} as the innermost node, and we call those
nodes farthest from the central node as outermost nodes. Then, the
T-graph can be alteratively constructed in another method, see
Fig.~\ref{copy}. Given the generation $g$, $\mathbb{T}_{g+1}$ may be
obtained by joining three copies of $\mathbb{T}_{g}$, denoted as
$\mathbb{T}_{g}^{(1)}$, $\mathbb{T}_{g}^{(2)}$, and
$\mathbb{T}_{g}^{(3)}$, respectively. In other words, to get
$\mathbb{T}_{g+1}$ one can merge together separate outermost nodes
of the three replicas of $\mathbb{T}_{g}$. In the process of
joining, the three outermost nodes are emerged into a single new
node, which becomes the innermost node of $\mathbb{T}_{g+1}$.

%%%%%%%%%%%%%%%%%%%%%%%%%%%%%%%%%%%%%%%%%%%%%%%%%%%%%%%%%
% Figure  3
%%%%%%%%%%%%%%%%%%%%%%%%%%%%%%%%%%%%%%%%%%%%%%%%%%%%%%%%%%
%%%%%%%%%%%%%%%%%%%%%%%%%%%%%%%%%%%%%%%%%%%%%%%%%%%%%%%%%%

\section{\label{sec03}Formulation of the problem}

In this section, we investigate a minimal model for discrete-time
random walks of a particle on the T-graph $\mathbb{T}_{g}$. At each
time step, the walker moves from its current location to any of its
nearest neighbors with equal probability. A key quantity
characterizing such a random walk is the first-passage time (FPT),
in terms of which many other quantities can be expressed. We are
interested in the mean first-passage time (MFPT) between two
distinct nodes, averaged over all pairs.

It is well-known that the random walks addressed here can be
described by the processes of Markov chains~\cite{KeSn76}, the
fundamental matrices corresponding to which can be used to express
the FPT between any pair of nodes. However, the fundamental matrix
method for computing the MFPT in $\mathbb{T}_{g}$ requires computing
the inversion of $N_{g}$ matrices of size $(N_{g}-1) \times
(N_{g}-1)$, making it prohibitively difficult to calculate the
quantity concerned for all but small networks.

To bring down the high computational demands the fundamental matrix
method makes, one can apply the method of the pseudoinverse of the
Laplacian matrix~\cite{BeGr03} for the graph $\mathbb{T}_{g}$ that
random walks are performed on, which allows to compute the FPT
between two arbitrary nodes directly from the network topology and
requires inversion of a single $N_{g} \times N_{g}$ matrix. The
elements $l_{ij}^{g}$ of the Laplacian matrix $\textbf{L}_g$ for
graph $\mathbb{T}_{g}$ are defined as follows: $l_{ij}^{g}=-1$ if
the pair of two different nodes $i$ and $j$ is connected by a link,
otherwise $l_{ij}^{g}=0$; while $l_{ii}^{(g)}=k_i$ (i.e., degree of
node $i$). Then, the pseudoinverse of the Laplacian matrix
$\textbf{L}_g$ is~\cite{RaMi71}
\begin{equation}\label{Pinverse01}
 \textbf{L}_g^\dagger=\left(\textbf{L}_g-\frac{\textbf{e}_g\textbf{e}_g^\top}{N_g}\right)^{-1}+\frac{\textbf{e}_g\textbf{e}_g^\top}{N_g}\,,
\end{equation}
where $\textbf{e}_g$ is the $N_g$-dimensional ``one" vector, i.e.,
$\textbf{e}_g=(1,1,\ldots,1)^\top$.

Let $F_{ij}(g)$ denote the FPT for a walker in the T-graph
$\mathbb{T}_{g}$, starting from node $i$, to arrive at node $j$ for
the first time. Since the graph $\mathbb{T}_{g}$ is connected, the
pseudoinverse matrix $\textbf{L}_g^\dagger$ is well defined and the
entries $l_{ij}^{\dagger,g}$ can be applied to express the FPT
$F_{ij}(g)$ as follows~\cite{CaAb08}:
\begin{equation}\label{Hitting01}
 F_{ij}(g)=\sum_{m=1}^{N_g}\left(l_{im}^{\dagger,g}-l_{ij}^{\dagger,g}-l_{jm}^{\dagger,g}+l_{jj}^{\dagger,g}\right)l_{mm}^{g}\,,
\end{equation}
where $l_{mm}^{g}$ is the $m$ entry of the diagonal of the Laplacian
matrix $\textbf{L}_g$ as defined above. Then the sum, $F_{\rm
tot}(g)$, for the FPT between two nodes over all node pairs in graph
$\mathbb{T}_{g}$ reads
\begin{equation}\label{Hitting02}
 F_{\rm tot}(g)=\sum_{i\neq j}\sum_{j=1}^{N_g}F_{ij}(g)\,,
\end{equation}
and the MFPT, $\langle F \rangle_g$, is
\begin{equation}\label{Hitting03}
 \langle F \rangle_g=\frac{F_{\rm tot}(g)}{N_g(N_g-1)}=\frac{1}{N_g(N_g-1)}\sum_{i\neq
 j}\sum_{j=1}^{N_g}F_{ij}(g)\,.
\end{equation}

\begin{figure}
\begin{center}
\includegraphics[width=0.85\linewidth,trim=40 30 20 0]{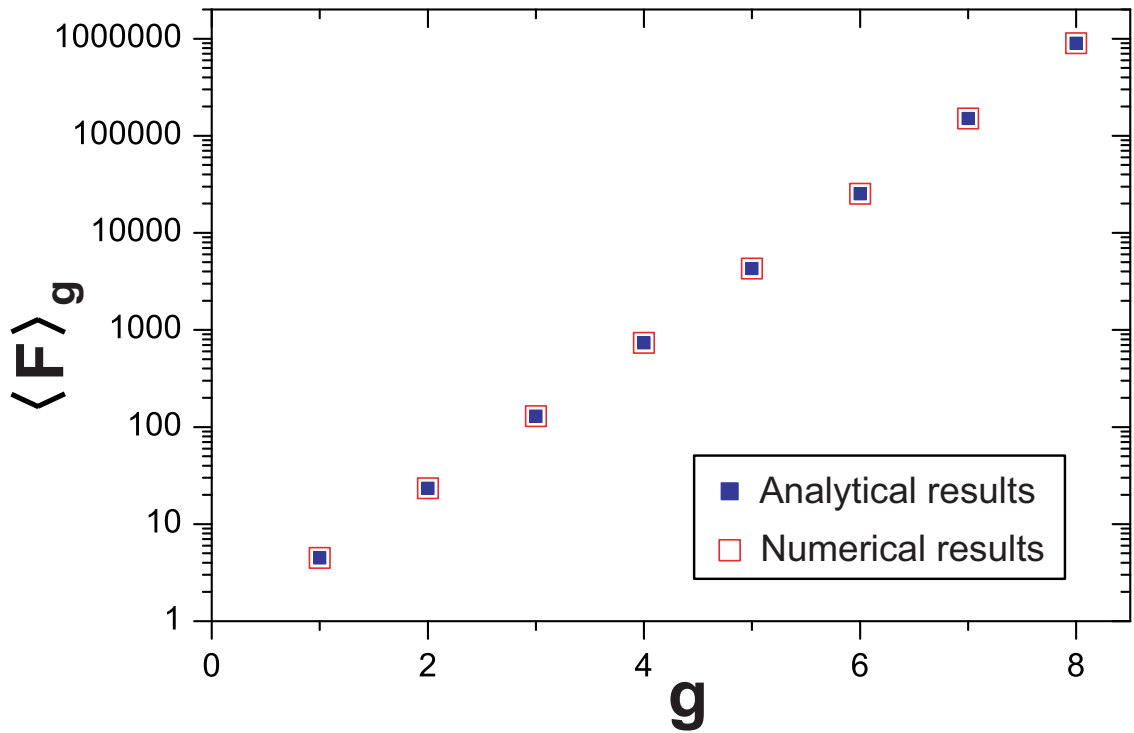}
\end{center}
\caption[kurzform]{\label{Time} (Color online) Mean first-passage
time $\langle F \rangle_g$ as a function of the iteration $g$ on a
semilogarithmic scale. The red squares are the numerical results
obtained by direct calculation from Eqs.~(\ref{Hitting01}) and
(\ref{Hitting03}), while the full blue squares represent the exact
values from Eq.~(\ref{Hitting07}), both of which agree with each
other.}
\end{figure}

The quantity of MFPT $\langle F \rangle_g$ is very important since
it measures the efficiency of the random walks on $\mathbb{T}_{g}$:
The smaller the value $\langle F \rangle_g$, the higher the
efficiency, and vice versa. Equations~(\ref{Hitting01}) and
(\ref{Hitting03}) show that the problem of calculating $\langle F
\rangle_g$ is reduced to determining the elements of the
pseudoinverse matrix $\textbf{L}_g^\dagger$, whose complexity is
that of inverting an $N_g \times N_g$ matrix, it can be easily
obtained by utilizing a standard software package, Mathematica 5.0.
However, since $N_g$ increases exponentially with $g$, for large
$g$, it becomes difficult to obtain $\langle F \rangle_g$ through
direct calculation using the pseudoinverse matrix, because of the
limitations of time and computer memory. Therefore, one can compute
directly the MFPT only for the first generations, see
Fig.~\ref{Time}. Fortunately, the particular construction of the
T-graph and the connection~\cite{ChRaRuSm89,Te91} between effective
resistance and first-passage time allow us to calculate analytically
the MFPT to obtain an explicit formula. Details will be given below.

%%%%%%%%%%%%%%%%%%%%%%%%%%%%%%%%%%%%%%%%%%%%%%%%%%%%%%%%%
% Figure 4
%%%%%%%%%%%%%%%%%%%%%%%%%%%%%%%%%%%%%%%%%%%%%%%%%%%%%%%%%%
%%%%%%%%%%%%%%%%%%%%%%%%%%%%%%%%%%%%%%%%%%%%%%%%%%%%%%%%%%

\section{Rigorous solution to mean first-passage time}

In the following text, we will use the connection between the
electrical networks and random walks to derive the closed-form
expression for MFPT $\langle F \rangle_g$, avoiding the computation
process for inverting any matrix.

\subsection{Relation for commute time and resistance distance between two nodes}

Given a graph, the underlying electrical network~\cite{DoSn84} can
be obtained by replacing each edge of the original graph with a unit
resistor. The effective resistance between any node pair $i$ and $j$
is defined as the voltage when a unit current enters one node and
leaves the other. It has been proved that the effective resistance
is a distance measure, so the quantity is also called resistance
distance~\cite{KlRa93}. Thus, we can view $\mathbb{T}_{g}$ as a
resistor network by considering all edges of $\mathbb{T}_{g}$ to be
unit resistors. Previous work~\cite{ChRaRuSm89,Te91} showed that
many problems on a resistor network are closed related to the
classic random walks on the original graph. For example, the
computation of effective resistance between two nodes in a resistor
network can be expressed by the FPT between the two nodes on the
corresponding network: For an arbitrary connected graph, the
effective resistance $R_{ij}$ between a pair of nodes $i$ and $j$ is
equal to $(F_{ij}+F_{ji})/(2\,E)$, where $E$ is the total number of
edges in the graph, and $F_{ij}$ is the expected time for a walker
starting at node $i$ to first reach $j$. In fact, $F_{ij}+F_{ji}$ is
the average time that a walker, originating at node $i$, will take
to hit node $j$ and go back to $i$, and it is often called commute
time~\cite{GoJa74} between $i$ and $j$ denoted as $C_{ij}$, i.e.,
$C_{ij}=F_{ij}+F_{ji}$.

According to the close relation between commute time and effective
resistance, we have that for $\mathbb{T}_{g}$, the effective
resistance $R_{ij}(g)$ between a pair of nodes $i$ and $j$ is
$R_{ij}(g)=C_{ij}(g)/(2\,E_{g})$, where
$C_{ij}(g)=F_{ij}(g)+F_{ji}(g)$ is the commute time between $i$ and
$j$. Thus $C_{ij}(g)=C_{ji}(g)=2\,E_{g}\,R_{ij}(g)$. Then,
Eq.~(\ref{Hitting02}) can be rewritten as
\begin{equation}\label{Hitting04}
 F_{\rm tot}(g)=\frac{1}{2}\sum_{i\neq j}\sum_{j=1}^{N_g}C_{ij}(g)=E_{g}\,\sum_{i\neq
 j}\sum_{j=1}^{N_g}R_{ij}(g)\,,
\end{equation}
and Eq.~(\ref{Hitting03}) can be recast as
\begin{equation}\label{Hitting05}
 \langle F \rangle_g=\frac{F_{\rm tot}(g)}{N_g(N_g-1)}=\frac{1}{N_g}\sum_{i\neq
 j}\sum_{j=1}^{N_g}R_{ij}(g)\,.
\end{equation}
Equation~(\ref{Hitting05}) shows that if we know how to determine
the effective resistances, then we have the way to find the MFPT.

For a general graph with order $N$, the complexity of resistance
computation is that of inverting an $N \times N$ matrix, identical
to that of calculating MFPT. However, for a treelike graph, the
effective resistance between two nodes is exactly the usual
shortest-path distance (also called geodesic distance) between the
corresponding nodes. Since the T-graph has a treelike structure and
it is self-similar at the same time, we can use these properties to
find its geodesic distance, obtaining a rigorous expression.

\subsection{Exact expression for mean first-passage time}

After reducing the determination of MFPT to finding the average
geodesic distance, the next step is to explicitly determine the
latter quantity. To this end, we represent all the shortest-path
distance of $\mathbb{T}_{g}$ as a matrix~\cite{Mietal92}, where the
entry $d_{ij}(g)$ is the shortest distance from node $i$ to node
$j$, which is the minimum length for the path connecting the two
nodes. The maximum value $D_{g}$ of $d_{ij}(g)$ is called the
diameter of $\mathbb{T}_{g}$. Then, the average geodesic distance of
$\mathbb{T}_{g}$ is defined as the mean of shortest-path distances
over all pairs of nodes:
\begin{equation}\label{apl01}
\langle L \rangle_{g} = \frac{S_g}{N_g(N_g-1)/2}\,,
\end{equation}
where
\begin{equation}\label{total01}
    S_g =\sum_{i \in \mathbb{T}_{g},\, j \in \mathbb{T}_{g},\, i \neq
    j}d_{ij}(g)
\end{equation}
denotes the sum of the geodesic distances between two nodes over all
pairs. Note that in Eq.~(\ref{total01}), for two nodes $i$ and $j$
($i \neq j$), we only count $d_{ij}(g)$ or $d_{ji}(g)$, not both.

We continue by exhibiting the procedure of the determination of the
total shortest-path distance and present the recurrence formula,
which allows us to obtain $S_{g+1}$ of the $g+1$ generation from
$S_{g}$ of the $g$ generation. According to the second construction
method, we can easily see that the T-graph is self-similar. This
obvious self-similar structure allows to calculate $S_g$
analytically. It is not difficult to find that the total
shortest-path distance $S_{g+1}$ satisfies the recursion
relation%~\cite{HiBe06,ZhZhZoChGu07,ZhChFaZhZhGu09}
\begin{equation}\label{total02}
  S_{g+1} = 3\, S_g + \Omega_g\,,
\end{equation}
where $\Omega_g$ is the sum over all shortest-path distances whose
end-points are not in the same $\mathbb{T}_{g}$ branch. The solution
of Eq.~(\ref{total02}) is
\begin{equation}\label{total03}
  S_g = 3^{g}\, S_0 + \sum_{m=0}^{g-1} \left[3^{g-m-1} \Omega_m\right].
\end{equation}
Thus, all that is left to obtain $S_g$ is to compute $\Omega_m$.

The paths that contribute to $\Omega_g$ must all go through the
innermost node (e.g., $X$ in Fig.~\ref{copy}) of $\mathbb{T}_{g+1}$.
To find $\Omega_g$, we denote $\Omega_g^{\alpha,\beta}$ as the sum
of all shortest paths with end-points in $\mathbb{T}_{g}^{(\alpha)}$
and $\mathbb{T}_{g}^{(\beta)}$, respectively. Notice that
$\Omega_g^{\alpha,\beta}$ rules out the paths where end-point is the
node shared by $\mathbb{T}_{g}^{(\alpha)}$ and
$\mathbb{T}_{g}^{(\beta)}$. By symmetry,
$\Omega_g^{1,2}=\Omega_g^{1,3}=\Omega_g^{2,3}$. Then, the total sum
$\Omega_g$ is given by
\begin{equation}\label{cross01}
\Omega_g
=\Omega_g^{1,2}+\Omega_g^{1,3}+\Omega_g^{2,3}=3\,\Omega_g^{1,2}\,.
\end{equation}

In order to calculate the path length $\Omega_g^{1,2}$, we give the
following notation. Let $s_{g}$ be the geodesic distance of a
outermost node of $\mathbb{T}_{g}$~\cite{EnJaSn76}, which is defined
as the sum of geodesic distances between the outermost node and all
nodes of $\mathbb{T}_{g}$, including the outermost node itself.
Prior to determining $s_{g}$, we first compute the diameter $D_{g}$
of $\mathbb{T}_{g}$. Actually, the diameter $D_{g}$ is equal to the
path length between arbitrary pair of the outermost nodes previously
belong to two different $D_{g-1}$ branches. Obviously, the following
recursive relation holds:
\begin{equation}\label{diam01}
D_{g+1} =2\,D_{g}\,.
\end{equation}
Considering the initial condition $D_{0}=1$, Eq.~(\ref{diam01}) is
solved inductively to obtain
\begin{equation}\label{diam02}
D_{g}=2^{g}\,.
\end{equation}

We now calculate the quantity $s_{g+1}$. Let $K$ denote an outermost
node of $\mathbb{T}_{g+1}$, which is in the branch
$\mathbb{T}_{g}^{(1)}$ (see Fig.~\ref{copy}). By definition,
$s_{g+1}$ can be given by the sum
\begin{eqnarray}\label{dist01}
s_{g+1}&=&\sum_{j\in \mathbb{T}_{g+1}} d_{Kj}(g+1)\nonumber \\
&=&\sum_{j \in \mathbb{T}_{g}^{(1)}} d_{Kj}(g+1)+\sum_{\stackrel{j
\in \mathbb{T}_{g}^{(2)}}{j \ne X}}d_{Kj}(g+1)+\sum_{\stackrel{j
\in \mathbb{T}_{g}^{(3)}}{j \ne X}}d_{Kj}(g+1)\nonumber \\
&=&s_g+2\,\sum_{\stackrel{j \in \mathbb{T}_{g}^{(2)}}{j \ne
X}}d_{Kj}(g+1)\,,
\end{eqnarray}
where we have made use of $\sum_{j \in \mathbb{T}_{g}^{(2)},\, j\neq
X}d_{Kj}(g+1)=\sum_{j \in \mathbb{T}_{g}^{(3)},\, j\neq
X}d_{Kj}(g+1)$, which is obvious from the symmetry. We denote the
second term in Eq.~(\ref{dist01}) by $2\,h_g$. Then,
$s_{g+1}=s_g+2\,h_g$. The quantity $h_g$ is evaluated as follows:
\begin{eqnarray}\label{dist02}
h_{g}&=&\sum_{\stackrel{j
\in \mathbb{T}_{g}^{(2)}}{j \ne X}}\Big[d_{KX}(g+1)+d_{Xj}(g+1)\Big]\nonumber \\
&=&s_g+(N_{g}-1)D_{g}\,,
\end{eqnarray}
where $d_{KX}(g+1)=D_{g}$ was used. Hence, Eq.~(\ref{dist01})
becomes
\begin{equation}\label{dist03}
s_{g+1}=3\,s_{g}+2\,(N_{g}-1)D_{g}.
\end{equation}
Using $N_g=3^{g}+1$, $D_{g} =2^{g}$, and $s_0=1$, Eq.~(\ref{dist03})
is resolved by induction
\begin{equation}\label{dist04}
s_{g}=3^{g-1}(2^{g+1}+1)\,.
\end{equation}

With above obtained results, we can determine $\Omega_g^{1,2}$,
which can be expressed in terms of the previously explicitly
determined quantities. By definition, $\Omega_g^{1,2}$ is given by
the sum
\begin{eqnarray}\label{cross02}
\Omega_g^{1,2}&=& \sum_{\stackrel{i \in \mathbb{T}_{g}^{(1)},\,i \ne
X}{j \in \mathbb{T}_{g}^{(2)},\, j \ne X}} d_{ij}(g+1)\nonumber \\
 &=&
\sum_{\stackrel{i \in \mathbb{T}_{g}^{(1)},\,i \ne X}{j \in
\mathbb{T}_{g}^{(2)},\, j \ne X}}
\Big[d_{iX}(g+1)+d_{Xj}(g+1)\Big]\nonumber \\
&=& (N_g-1)\,\sum_{\stackrel{i \in \mathbb{T}_{g}^{(1)}}{i \ne X}}
d_{iX}(g+1)+(N_g-1)\,\sum_{\stackrel{j \in \mathbb{T}_{g}^{(2)}}{j
\ne X}} d_{Xj}(g+1)\nonumber \\&=& 2\,(N_g-1)\,s_g \nonumber
\\&=&2\times3^{2g-1}(2^{g+1}+1)\,,
\end{eqnarray}
where we have used the equivalence relation $\sum_{i \in
\mathbb{T}_{g}^{(1)},\, i \ne X} d_{iX}(g+1)=\sum_{j\in
\mathbb{T}_{g}^{(2)},\,j \ne X}d_{Xj}(g+1)$\,.

Inserting Eq.~(\ref{cross02}) into Eq.~(\ref{cross01}), we have
\begin{equation}\label{cross04}
\Omega_g=2\times3^{2g}(2^{g+1}+1)\,.
\end{equation}
Substituting Eq.~(\ref{cross04}) into Eq.~(\ref{total03}) and using the initial value $S_0=1$, %Eq.~(\ref{total04}) can be resolved inductively to
we can obtain the exact expression for the total distance
\begin{equation}\label{Sg}
S_g=\frac{1}{5}\times3^{g-1}(4\times 6^{g}+5\times 3^{g}+6).
\end{equation}
Then the analytic expression for mean geodesic distance can be
obtained as
\begin{equation}\label{apl02}
\langle L \rangle _g =\frac{8\times 6^{g}+10\times3^{g}+12}{15
\times (3^{g}+1)}\,.
\end{equation}

Since for $\mathbb{T}_{g}$, the shortest-path distance is equivalent
to the resistance distance, i.e., $d_{ij}(g)=R_{ij}(g)$, according
to the above obtained results shown in Eqs.~(\ref{Sg})
and~(\ref{apl02}), we can easily get that the total of FPT between
all $N_g(N_g-1)$ pairs of nodes and the MFPT are
\begin{equation}\label{Hitting06}
 F_{\rm tot}(g)=2\,E_{g}\,S_g=\frac{2}{15}\times 9^{g}(4\times 6^{g}+5\times 3^{g}+6)
\end{equation}
and
\begin{equation}\label{Hitting07}
 \langle F \rangle_g=\frac{2\,S_{g}}{N_g}=E_g\langle L \rangle _g=\frac{3^{g}}{15(3^{g}+1)}(8\times 6^{g}+10\times3^{g}+12)\,,
\end{equation}
respectively.

We have checked our analytical formula against numerical values, see
Fig.~\ref{Time}. For the range of $1 \leq g \leq 8$, the values
obtained from Eq.~(\ref{Hitting07}) are in complete agreement with
those numerical results on the basis of the direct calculation
through the method of pseudoinverse matrix discussed in
section~\ref{sec03}. This agreement serves as an independent test of
our theoretical formula.

Note that the result provided in Eq.~(\ref{Hitting07}) is consistent
with the recent proposal by Condamin \emph{et
al.}~\cite{CoBeTeVoKl07} for a very general scaling form for FPT,
$\langle F \rangle$, of a random walk as a function of the distance,
$L$, between the origin and the trap location. Condamin \emph{et
al.} proved that, for complex invariant networks with $d_w>d_f$, the
leading behavior of FPT behaves as $\langle F \rangle \sim
NL^{d_w-d_f}$. For the T-graph, $d_w-d_f=1$, according to the
conclusion of Condamin \emph{et al.}, one has $\langle F \rangle_g
\sim {N_g}\langle L \rangle _g$ as found in Eq.~(\ref{Hitting07}).

Next we will express the MFPT $\langle F \rangle_g$ as a function of
network order $N_g$, in order to obtain the scaling between these
two quantities. From the relation $N_g=3^g+1$, we have $3^{g}=N_g-1$
and $g=\log_3\big(N_g-1\big)$. Thereby, we can rewrite
Eq.~(\ref{Hitting07}) as
\begin{equation}\label{Hitting08}
\langle F \rangle_g =
\frac{2(N_g-1)}{15N_g}\Big(4(N_g-1)^{1+\log_32} +5(N_g-1)+6\Big) \,,
\end{equation}
which provides explicit dependence of $\langle F \rangle_g$ on
network order $N_g$. In the infinite network order, namely $N_g
\rightarrow \infty$,
\begin{equation}\label{Hitting09}
\langle F \rangle_g \approx
\frac{8}{15}(N_g)^{1+\log_32}=\frac{8}{15}(N_g)^{2/\tilde{d}}\,.
\end{equation}
Thus, for the large T-graph, the MFPT grows as a power law function
of network order with the exponent larger than 1 and less than 2,
implying that the MFPT increases superlinearly with the number of
network nodes.

Equation~(\ref{Hitting09}) encodes the speed of the random walks on
$\mathbb{T}_{g}$, which may be quantified by the coverage $C_g(t)$,
standing for the mean number of different nodes visited by a walker
at time $t$, averaged for distinct walks initially starting from
different sources~\cite{BlZuKl84,SzZwAg88}. As $t \rightarrow
\infty$, the leading asymptotic behavior for $C_g(t)$ is
\begin{equation}\label{coverage01}
C_g(t) \sim t^{\tilde{d}/2}\,,
\end{equation}
which may be established from the following heuristic
argument~\cite{KaBa02PRE,KaBa02IJBC}. Since the random walker will
visit all nodes of $\mathbb{T}_{g}$ with probability one in the
longtime limit, irrespective of its starting location, $N_g$ plays
the role of $C_g(t)$, and $\langle F \rangle_g$ is like $t$ (step
number of the walker). Then, inverting the relation in
Eq.~(\ref{coverage01}) leads to asymptotic scaling $\langle F
\rangle_g \sim (N_g)^{2/\tilde{d}}$, as found above and given by
Eq.~(\ref{Hitting09}), which also provides the exact coefficient of
proportionality.

Notice that in Ref.~\cite{Ag08} Agliari studied the trapping problem
with an immobile trap located at the innermost (central) node, and
showed that in the asymptotic limit, the average trapping time (ATT)
to first reach the trap, averaged over all nodes except the
absorbing node itself, exhibiting a similar behavior (the same
exponent) as that of Eq.~(\ref{Hitting09}), but has a different
prefactor. Note that, due to the symmetry, the ATT in this case
corresponds to the mean time for random walks with a fixed trap
situated at an outmost node in the T-graph of generation
$g-1$~\cite{Ag08}. That is to say, random walks of these two extreme
cases show almost the same behavior of ATT. On the other hand, the
random walks discussed here may be considered as the trapping issue
with the trap uniformly distributed throughout all nodes on the
T-graph. Thus, we can conclude that the trap's location has no
qualitative impact on the leading behavior of the ATT for random
walks on the T-graph, which is in sharp contrast to trapping problem
on some scale-free networks~\cite{KiCaHaAr08}, where the ATT depends
on the trap position: it is much smaller for the case when trap is
located at a large-degree node than that for the case when the trap
is uniformly placed.

\section{Conclusions}

In this paper, we have studied the standard discrete-time random
walks on the T-graph. By making use of the link between the electric
networks and random walks, we have studied analytically the MFPT
averaged over all pairs of nodes, based on the recursive relations
derived from the self-similar structure of the T-graph. We have
determined the rigorous solution for the MFPT, which in the large
limit of graph order, increases algebraically with the number of
nodes. We showed that trap location has little influence on the
scaling of MFPT for random walks on the T-graph. We expect that by
providing a paradigm for computing the MFPT, our analytical
technique could guide and shed light on related studies of random
walks on other deterministic media, especially to some scale-free
fractals~\cite{SoHaMa05,SoHaMa06} that has received much recent
contention~\cite{ZhZhZo07,ZhZhChGu08}. We also expect that our work,
especially the exact solution, could prompt the studies of diffusion
process on random fractals, other stochastic graphs, and even
complex networks~\cite{AlBa02}, by giving a guide to and a test of
approximate methods for random media.

\section*{Acknowledgment}

This research was supported by the National Basic Research Program
of China under grant No. 2007CB310806, the National Natural Science
Foundation of China under Grant Nos. 60704044, 60873040 and
60873070, Shanghai Leading Academic Discipline Project No. B114, and
the Program for New Century Excellent Talents in University of China
(NCET-06-0376).


\section*{References}
\begin{thebibliography}{10}



\bibitem{Ei1906}
A. Einstein, Ann. Phys. (Leipzig) {\bf 17}, 549 (1905); {\bf 19},
371 (1906).

\bibitem{Sm1906}
M. Smoluchowski, Ann. Phys. (Leipzig) {\bf 21}, 756 (1906).

\bibitem{Sp1964}
F. Spitzer, \emph{Principles of Random Walk} (Van Nostrand,
Princeton, New Jersey, 1964).

\bibitem{We1994}
G. H. Weiss, \emph{Aspects and Applications of the Random Walk}
(North Holland, Amsterdam, 1994).

\bibitem{Hu1996}
B. D. Hughes, \emph{Random Walks and Random Environments: Random
Walks} (Clarendon Press, Oxford, 1996), Vol. 1.

\bibitem{MeKl00}
R. Metzler and J. Klafter, Phys. Rep. {\bf 339}, 1 (2000).

\bibitem{BeHa00}
D. ben-Avraham and  S. Havlin, \emph{Diffusion and Reactions in
Fractals and Disordered Media} (Cambridge Universiy Press,
Cambridge, 2000).

\bibitem{HaBe02}
S. Havlin and D. ben-Avraham, Adv. Phys. {\bf 51}, 187 (2002).

\bibitem{MeKl04}
R. Metzler and J. Klafter, J. Phys. A: Math. Gen. {\bf 37}, R161
(2004).

\bibitem{BuCa05}
R Burioni and D Cassi, J. Phys. A: Math. Gen. {\bf 38}, R45 (2005).


\bibitem{Ma82}
B. Mandlebrot, \emph{The Fractal Geometry of Nature} (Freeman, San
Francisco, 1982).

\bibitem{Vi83}
T. Vicsek J. Phys. A: Math. Gen. {\bf 16}, L647 (1983).

\bibitem{ZhZhChYiGu08}
Z. Z. Zhang, S. G. Zhou, L. C. Chen, M. Yin, and J. H. Guan, J.
Phys. A: Math. Theor. {\bf 41}, 485102 (2008).

\bibitem{BeOs79}
A. N. Berker and S. Ostlund, J. Phys. C {\bf 12}, 4961 (1979).

\bibitem{GrKa82}
R. Griffiths and M. Kaufman, Phys. Rev. B {\bf 26}, 5022 (1982).

\bibitem{ZhZhFaGuZh07}
Z.Z. Zhang, S. G. Zhou, L. J. Fang, J. H. Guan, and Y. C. Zhang, EPL
{\bf 79}, 38007 (2007).

\bibitem{ZhZhZoCh08}
Z. Z. Zhang, S. G. Zhou, T. Zou, and G. S. Chen, J. Stat. Mech.
P09008 (2008).

\bibitem{KaRe86}
S. Havlin and H. Weissman, J. Phys. A: Math. Gen. {\bf 19}, L1021
(1986).

\bibitem{KaRe89}
B. Kahng and S. Redner, J. Phys. A: Math. Gen. {\bf 22}, 887 (1989).

\bibitem{Ma89}
A. Maritan, Phys. Rev. Lett. {\bf 62}, 2845 (1989).

\bibitem{MaSaSt93}
A. Maritan, G. Sartoni, and A. L. Stella, Phys. Rev. Lett. {\bf 71},
1027 (1993).

\bibitem{GiMaNat94}
A. Giacometti, A. Maritan, and H. Nakanishi, J. Stat. Phys. {\bf
75}, 669 (1994).

\bibitem{KnKn96}
M. Kne\u{z}evi\'{c} and D. Kne\u{z}evi\'{c}, Phys. Rev. E {\bf 83},
2130 (1996).

\bibitem{MaHa89}
O. Matan and S. Havlin, Phys. Rev. A {\bf 40}, 6573 (1989).

\bibitem{BeMeTeVo08}
O. B\'enichou, B. Meyer, V. Tejedor, and R. Voituriez, Phys. Rev.
Lett. {\bf 101}, 130601 (2008).

\bibitem{HaRo08}
C. P. Haynes and A. P. Roberts, Phys. Rev. E {\bf 78}, 041111
(2008).

\bibitem{Ag08}
E. Agliari, Phys. Rev. E {\bf 77}, 011128 (2008).

\bibitem{KiCaHaAr08}
A. Kittas, S. Carmi, S. Havlin, and P. Argyrakis, EPL {\bf 84},
40008 (2008).

\bibitem{DoSn84}
P. G. Doyle and J. L. Snell, \emph{Random Walks and Electric
Networks} (The Mathematical Association of America, Oberlin, OH,
1984); e-print arXiv:math.PR/0001057.


\bibitem{Ge82}
P. G. de Gennes, J. Chem. Phys. {\bf 76}, 3316 (1982).


\bibitem{KeSn76}
J. G. Kemeny and J. L. Snell, \emph{Finite Markov Chains} (Springer,
New York, 1976).

\bibitem{BeGr03}
A. Ben-Israel and T. Greville, \emph{Generalized Inverses: Theory
and Applications}, 2nd ed. (Springer, New York, 2003).

\bibitem{RaMi71}
C. Rao and S. Mitra, \emph{Generalized Inverse of Matrices and Its
Applications} (John Wiley and Sons, 1971).

\bibitem{CaAb08}
A. G. Cant\'u and E. Abad, Phys. Rev. E {\bf 77}, 031121 (2008).

\bibitem{ChRaRuSm89}
A. K. Chandra, P. Raghavan, W. L. Ruzzo, and R. Smolensky, in
\emph{Proceedings of the 21st Annnual ACM Symposium on the Theory of
Computing} (ACM Press, New York, 1989), pp. 574-586.

\bibitem{Te91}
P. Tetali, J. Theor. Probab. {\bf 4}, 101 (1991).


\bibitem{KlRa93}
D. J. Klein and M. Randi\'c, J. Math. Chem. {\bf 12}, 81 (1993).

\bibitem{GoJa74}
F. Gobel and A. Jagers, Stoch. Proc. Appl. {\bf 2}, 311 (1974).

\bibitem{Mietal92}
Z. Mihali\'c, D. Veljan, D. Ami\'c, S. Nikoli\'c, D. Plavsi\'c, and
N. Trinajsti\'c, J. Math. Chem. {\bf 11}, 223 (1992).


\bibitem{EnJaSn76}
R. C. Entringer, D. E. Jackson, and D. A. Snyder, Czechoslovak Math.
J. {\bf 26} 283 (1976).


\bibitem{CoBeTeVoKl07}
S. Condamin, O. B\'enichou, V. Tejedor, R. Voituriez, and J.
Klafter, Nature (London) {\bf 450}, 77 (2007).

\bibitem{BlZuKl84}
A. Blumen, G. Zumofen, and J. Klafter, Phys. Rev. B {\bf 30}, 5379
(1984).

\bibitem{SzZwAg88}
A. Szabo, R. Zwanzig, and N. Agmon, Phys. Rev. Lett. {\bf 61}, 2496
(1988).

\bibitem{KaBa02PRE}
J. J. Kozak and V. Balakrishnan, Phys. Rev. E {\bf 65}, 021105
(2002).

\bibitem{KaBa02IJBC}
J. J. Kozak and V. Balakrishnan, Int. J. Bifurcation Chaos Appl.
Sci. Eng. {\bf 12}, 2379 (2002).

\bibitem{SoHaMa05}
 C. Song, S. Havlin, H. A. Makse,
Nature (London) {\bf 433}, 392 (2005).

\bibitem{SoHaMa06}
 C. Song, S. Havlin, H. A. Makse,
Nature Phys. {\bf 2}, 275 (2006).

\bibitem{ZhZhZo07}
Z. Z. Zhang, S. G. Zhou, and T. Zou, Eur. Phys. J. B {\bf 56}, 259
(2007).

\bibitem{ZhZhChGu08}
Z. Z. Zhang, S. G. Zhou, L. C. Chen, and J. H. Guan, Eur. Phys. J. B
{\bf 64}, 277 (2008).

\bibitem{AlBa02}
R. Albert and A.-L. Barab\'asi,
      %Statistical mechanics of complex networks,
       Rev. Mod. Phys. {\bf 74}, 47 (2002).

\end{thebibliography}
\end{document}